\documentclass{emulateapj}
\usepackage{apjfonts}

\newcommand{\HII}{H\,{\sc ii}}
\newcommand{\mjb}{mJy~beam$^{-1}$}

\newcommand{\et}{et~al.}

\newcommand{\mum}{$\mu$m}

\shortauthors{Brogan \et\/}  \shorttitle{35 New SNRs}
\begin{document}

\title{Discovery of 35 New Supernova Remnants in the Inner Galaxy}

\author{C.~L. Brogan\altaffilmark{1}, J. D. Gelfand\altaffilmark{2},
B. M. Gaensler\altaffilmark{2}, N.~E.~Kassim\altaffilmark{3},
T.~J.~Lazio\altaffilmark{3}}

\altaffiltext{1}{Institute for Astronomy, 640 North A`ohoku Place, Hilo, HI 
96720; cbrogan@ifa.hawaii.edu.}

\altaffiltext{2}{Harvard-Smithsonian Center for Astrophysics, 
60 Garden Street, Cambridge, MA 02138}

\altaffiltext{3}{Remote Sensing Division, Naval Research Laboratory,
Washington DC 20375-5351}

\begin{abstract}

We report the discovery of up to 35 new supernova remnants (SNRs) from a
$42\arcsec$ resolution 90cm multi-configuration Very Large Array
survey of the Galactic plane covering $4.5\arcdeg<\ell <22.0\arcdeg$ and
$\mid b\mid < 1.25\arcdeg$. Archival 20cm, 11cm, and 8 \mum\/ data
have also been used to identify the SNRs and constrain their
properties. The 90cm image is sensitive to SNRs with diameters
$2.5\arcmin$ to $50\arcmin$ and down to a surface brightness limit
of $\sim 10^{-21}$ W~m$^{-2}$~Hz$^{-1}$~sr$^{-1}$.  This survey has
nearly tripled the number of SNRs known in this part of the Galaxy,
and represents an overall 15\% increase in the total number of
Galactic SNRs.  These results suggest that further deep low frequency
surveys of the inner Galaxy will solve the discrepancy between the
expected number of Galactic SNRs and the significantly smaller number
of currently known SNRs.

\end{abstract}


\section{INTRODUCTION}

Statistical studies of supernova (SN) rates, based on OB star counts,
pulsar birth rates, Fe abundance, and the SN rate in other Local Group
galaxies, suggest that there should be many more supernova remnants
(SNRs) in our Galaxy \citep[$\ga
1000$;][]{Li1991,Tammann1994} than are currently known
\citep[$\sim 231$;][]{Green2004}.  This deficit is likely the result
of selection effects acting against the discovery of old, faint, large
remnants, as well as young, small remnants in previous low resolution
and/or poor sensitivity Galactic radio surveys (e.g.\ Green 1991).
For example, past single dish surveys with $\lambda\ge 11$cm have
resolutions $>4\arcmin$ \citep[e.g.][]{Haslam1982,Reich1990}.  The
20cm NRAO Very Large Array (VLA) Sky Survey (NVSS) has high resolution
($\sim 45\arcsec$) but poor surface brightness sensitivity
\citep[][]{Condon1998}. The missing remnants are likely concentrated
toward the inner Galaxy where the diffuse Galactic plane synchrotron
emission and thermal \HII\/ regions cause the most confusion.  This
premise is supported by the large number of new SNRs discovered in the
4th quadrant by \citet{Whiteoak1996} using the Molongolo Observatory
Synthesis Telescope (MOST) at 35cm with $\sim 45\arcsec$ resolution
and comparatively high sensitivity. Despite the success of the MOST
survey, a significant shortfall compared to predictions remains --
most likely due to limited dynamic range.

Thus, more sensitive, high resolution surveys of the inner Galaxy at
low radio frequencies are the key to determining whether the
``missing'' remnants exist or if our understanding of SN rates is
significantly flawed. From multi-configuration VLA 90cm (330 MHz)
observations of just 1~deg${}^2$ centered on SNR G11.2-0.3, Brogan et
al. (2004) discovered three new SNRs, demonstrating the power of such
observations.  This Letter presents the discovery of 35 SNRs
(including the three described above) from a more extensive 90cm VLA
survey.

\section{OBSERVATIONS AND RESULTS}

We have imaged the Galactic plane at 90cm from $\ell=+4.5\arcdeg$ to
$+22\arcdeg$ and $\mid b\mid < 1.25\arcdeg$ using the VLA in the B, C,
and D configurations between 2002 and 2004. The VLA 90cm FWHM primary
beam is $2.5\arcdeg$ and the mosaiced image is comprised of 14
pointings, each separated by $\sim 1.25\arcdeg$. The majority of these
data are new, but VLA archival data are also included for regions near
$\ell=6\arcdeg$, $9\arcdeg$, $11\arcdeg$, \& $21\arcdeg$. Each
pointing was observed for approximately 1.2, 1.6, and 2.2 hours in the
D, C, and B configurations, respectively, over a wide range of hour
angles.  The data were reduced using AIPS and standard wide-field, low
frequency data reduction techniques\footnote{Described at {\em
http://www.vla.nrao.edu/astro/guides/p-band/}}
\citep[e.g.][]{Brogan2004}.

In order to achieve high resolution images that are also sensitive to
extended structure, we have employed the multi-scale (MS) cleaning
algorithm in AIPS \cite[e.g.][]{Wakker1988}. Instead of a point source
clean model, MS clean tapers the data and cleans on a variety of
scales; this process virtually eliminates artifacts due to the ``clean
instability''. Three scales were used to create the 90cm mosaic: the
intrinsic resolution of the multi-configuration data of $\sim
40\arcsec\times 30\arcsec$ (differing slightly for each pointing),
$100\arcsec$, and $220\arcsec$. After cleaning, the individual images
were convolved to a common resolution of $42\arcsec$, corrected for
primary beam attenuation, and then linearly mosaiced.  These data are
not sensitive to structures $\ga 50\arcmin$, and thus resolve out the
Galactic background synchrotron emission.  The final rms noise of the
mosaiced image is a function of $b$ and is $\sim 5$ \mjb\/ for $\mid
b\mid\lesssim 0.5\arcdeg$, increasing to $\sim 8$ \mjb\/ at $\mid
b\mid=1\arcdeg$.  This is by far the highest dynamic range large-scale 
image \footnote{Image is available from C. Brogan upon request.} of this 
part of the Galactic plane yet created for $\lambda >
20$cm.

Complementary images at other frequencies are required to search for
non-thermal emission. For this purpose we have created a 20cm mosaic
of a large fraction of the 90cm survey region using 
data from the northern extension of the Southern Galactic
Plane Survey (SGPS) observed with the Australia Telescope Compact
Array \citep[see][survey region includes
$5\arcdeg<\ell<20\arcdeg$]{MG2005} combined with archival VLA
D-configuration 20cm data \citep[see][outside the SGPS region the VLA
data alone were used]{Helfand2005b}.  The resolution of this image is
$70\arcsec\times 37\arcsec$ and the rms noise is $\sim 15$ \mjb\/.
The resulting 20cm image is not sensitive to smooth structures larger
than about $\sim 18\arcmin$, for this reason the $4\farcm3$
resolution single dish Bonn 11cm survey data with an rms noise of
$\sim 8$ \mjb\/ were also utilized \citep{Reich1990}. {\em Midcourse
Space Experiment} (MSX) 8 \mum\/ data \citep{Price2001} with
$20\arcsec$ resolution were also used to distinguish between thermal 
and non-thermal emission (non-thermal emission is anti-correlated with 
bright IR emission).

\subsection{SNR Candidates}

The SNR candidates were selected based on the following criteria: 
\begin{enumerate}
\item The source must be resolved in our $42\arcsec$ 90cm image and show a
shell-like (or partial shell) morphology 

\item The radio continuum
spectral index ($S_{\nu}\propto\nu^{\alpha}$) computed from the
integrated flux densities must be negative, indicative of non-thermal
emission

\item The source must be distinct from bright mid-IR 8
\mum\/ emission. 
\end{enumerate}
Table~1 lists the 35 sources that meet these
criteria.  By way of confirmation, this list includes seven sources
that have been previously identified as SNR candidates that have not
yet been included in Green's SNR catalog \citep{Green2004}. The 19
previously identified SNRs within the survey region are also detected
and confirmed \citep[see][]{Green2004}.

For the determination of criterion
two both the 20 and 11cm flux density estimates suffer from
significant caveats. The 20cm flux densities ($S_{\rm 20 cm}$) for
candidates larger than about $10\arcmin$ are likely to be lower limits
(due to missing short spacing information); $S_{\rm 20 cm}$
measurements for candidates larger than $\sim 18\arcmin$ are
precluded. The 11cm flux densities suffer to varying degrees from
confusion due to the comparably low resolution of this survey. The
flux densities at all three wavelengths have been corrected for
background emission, though this correction is least certain for the
11cm data.  For the majority of candidates, the spectral indices
($\alpha$) determined from the 90/20 cm data and 90/11 cm data (last
two columns of Table~1) are in good agreement. 

\begin{deluxetable*}{rlcccccrrrccl}
\tabletypesize{\scriptsize}
\setlength{\tabcolsep}{0.02in}
\tablewidth{0pt}
\tablecaption{Properties of SNR Candidates\label{tab1}}
\tablecolumns{13}
\tablehead{
\colhead{$\ell$} & \colhead{$b$} & \colhead{R. A.} & \colhead{Dec.} &
\colhead{Size} & \colhead{Morphology$^a$} & \colhead{Class} & 
\colhead{$S_{\rm 90cm}$$^b$} & \colhead{$S_{\rm 20cm}$$^b$} & \colhead{$S_{\rm 11cm}$$^b$} & 
\colhead{$\alpha_{(90/20)}$$^c$} & \colhead{$\alpha_{(90/11)}$$^c$} & \colhead{Notes$^d$}\\
\colhead{($\arcdeg$)} & \colhead{($\arcdeg$)} & \colhead{(h m s)} & \colhead{($\arcdeg$~$\arcmin$)} &
\colhead{($\arcmin$x $\arcmin$)} & \colhead{} & \colhead{} & 
\colhead{(Jy)} & \colhead{(Jy)} & \colhead{(Jy)} & 
\colhead{} & \colhead{}  & \colhead{}}
\startdata
  5.55 & +0.32 & 17 57 04  & $-$24 00  & 12 x 15   & Shell & II & 14.3 (0.3) & 4.6 (0.9) & 3.9 (0.4) & $-$0.8 & $-$0.6 &  \\
  5.71 & $-$0.08 & 17 58 49  & $-$24 03  & 9 x 12  & Partial Shell & III & 4.3 (0.2) & 2.0 (0.6) & 1.6 (0.2) & $-$0.5 & $-$0.5 &  \\
  6.10 & +0.53 & 17 57 29  & $-$23 25  & 18 x 12 & Partial Shell & I & 13.4 (0.2) & 3.5 (0.7) & 2.0 (0.2) & $-$0.9 & $-$0.9 &  \\
  6.31 & +0.54 & 17 57 54  & $-$23 14  & 3 x 9 & Filament & III & 1.4 (0.1) & 0.7 (0.3) & 0.3 (0.1) & $-$0.5 & $-$0.8 &  \\
  6.51 & $-$0.48 & 18 02 11 & $-$23 34  & 18 x 18 & Shell & I & 60.8 (0.4) & 22.1 (1.1) & 22.7 (2.3) & $-$0.7 & $-$0.5 & H/O \\
  7.20 & +0.20 &  18 01 07  & $-$22 38  & 12 x 12  & Partial Shell & II & 5.2 (0.2) & 2.3 (0.7) & 1.8 (0.2) & $-$0.6 & $-$0.5 & H \\
  8.31 & $-$0.09 & 18 04 34 & $-$21 49 & 5 x 4 & Shell & II & 2.3 (0.1) & 1.0 (0.2) & 0.5 (0.1) & $-$0.6 & $-$0.7 &  H \\
  8.90 & +0.40 &  18 03 58  & $-$21 03  & 24 x 24 & Shell & II & 18.2 (0.5) & \nodata & 4.9 (0.5) & \nodata & $-$0.6 &  \\
  9.70 & $-$0.06 &  18 07 22  & $-$20 35  & 15 x 11 & Shell & I & 6.5 (0.2) & 3.0 (0.7) & 1.5 (0.2) & $-$0.5 & $-$0.7 & H \\
  9.95 & $-$0.81 &  18 10 41  & $-$20 43  & 12 x 12 & Shell & II & 11.0 (0.3) & 5.9 (0.8) & 4.6 (0.5) & $-$0.4 & $-$0.4 &  \\
 10.59 & $-$0.04 & 18 09 08 & $-$19 47 & 6 x 6 & Partial Shell & II & 1.4 (0.1) & 0.7 (0.3) & 0.3 (0.1) & $-$0.5 & $-$0.7 & X\\
 11.03 & $-$0.05 & 18 10 04 & $-$19 25 & 9 x 11 & Partial Shell & I & 3.1 (0.2) & 1.1 (0.5) & 1.0 (0.1) & $-$0.7 & $-$0.5 & O/X \\
 11.15 & $-$0.71 & 18 12 46 & $-$19 38 & 11 x 7 & Partial Shell & I & 2.3 (0.1) & 0.8 (0.4) & 0.4 (0.1) & $-$0.7 & $-$0.8 & H/O \\
 11.17 & $-$1.04 &  18 14 03  & $-$19 46  & 18 x 12 & Shell & I & 11.0 (0.3) & 4.7 (0.8) & 4.1 (0.4) & $-$0.6 & $-$0.5 &  O \\
 11.18 & +0.11 & 18 09 47 & $-$19 12  & 12 x 10 & Shell & I & 3.5 (0.2) & 2.0 (0.5) & 1.6 (0.2) & $-$0.4 & $-$0.4 &  H/O \\
 11.89 & $-$0.21 &  18 12 25  & $-$18 44  & 4 x 4 & Shell & II & 0.9 (0.1) & 0.6 (0.2) & 0.4 (0.1) & $-$0.3 & $-$0.4 & H/X\\
 12.26 & +0.30 &  18 11 17  & $-$18 10  & 5 x 6 & Partial Shell & I & 1.5 (0.1) & 0.6 (0.3) & 0.4 (0.1) & $-$0.7 & $-$0.6 & H \\
 12.58 & +0.22 & 18 12 14  & $-$17 55  & 5 x 6 & Composite? & II & 0.8 (0.1) & 0.5 (0.3) & 0.3 (0.1) & $-$0.4 & $-$0.5 &  \\
 12.72 & $-$0.00 &  18 13 19  & $-$17 54  & 6 x 6 & Shell & I & 2.0 (0.1) & 0.6 (0.3) & 0.3 (0.1) & $-$0.8 & $-$0.8 & H \\
 12.83 & $-$0.02 &  18 13 37  & $-$17 49  & 3 x 3 & Shell & I & 1.2 (0.1) & 0.7 (0.2) & 0.4 (0.1) & $-$0.4 & $-$0.5 & H/O/X \\
 14.18 & $-$0.12 &  18 15 52  & $-$16 34 & 6 x 5 & Shell & II & 0.9 (0.1) & 0.4 (0.3) & 0.3 (0.1) & $-$0.6 & $-$0.5 &  \\
 14.30 & +0.14 &  18 15 58  & $-$16 27  & 5 x 4 & Partial Shell & II & 1.2 (0.1) & 0.5 (0.3) & 0.6 (0.1) & $-$0.5 & $-$0.3 &  \\
 15.42 & +0.18 &  18 18 02  & $-$15 27  & 14 x 15 & Shell & I & 10.9 (0.3) & 4.6 (0.8) & 2.9 (0.3) & $-$0.6 & $-$0.6&  \\
 15.51 & $-$0.15 &  18 19 25  & $-$15 32  & 8 x 9 & Shell & III & 4.2 (0.2) & 1.9 (0.5) & 1.0 (0.1) & $-$0.5 & $-$0.7&  \\
 16.05 & $-$0.57 &  18 21 56  & $-$15 14  & 15 x 10 & Shell & I & 4.9 (0.2) & 2.2 (0.7) & 1.3 (0.1) & $-$0.6 & $-$0.6&  \\
 16.41 & $-$0.55 &  18 22 38  & $-$14 55  & 13 x 13 & Partial Shell & II & 10.0 (0.3) & 3.6 (0.9) & 2.0 (0.2) & $-$0.7 & $-$0.8 &  \\
 17.02 & $-$0.04 & 18 21 57  & $-$14 08  & 5 x 5 & Shell & I & 0.7 (0.1) & 0.4 (0.3) & 0.3 (0.1) & $-$0.5 & $-$0.4&  H \\
 17.48 & $-$0.12 & 18 23 08 & $-$13 46 & 6 x 6 & Partial Shell & II & 0.9 (0.1) & 0.3 (0.4) & 0.3 (0.1) & $-$0.7 & $-$0.6 &  \\
 18.16 & $-$0.16 &  18 24 34  & $-$13 11  & 8 x 8 & Shell & I & 7.6 (0.1) & 3.9 (0.4) & 3.0 (0.3) & $-$0.5 & $-$0.4 & H/O/X \\
 18.62 & $-$0.28 &  18 25 55  & $-$12 50  & 6 x 6 & Partial Shell & II & 1.9 (0.1) & 1.2 (0.4) & 0.7 (0.1) & $-$0.3 & $-$0.5 & H \\
 19.13 & +0.90 &  18 22 37  & $-$11 50  & 24 x 18 & Partial Shell & III & 27.2 (0.5) & \nodata & 9.4 (0.9) & \nodata & $-$0.5 &  \\
 19.15 & +0.27 & 18 24 56  & $-$12 07  & 27 x 27 & Partial Shell & II & 17.4 (0.4) & \nodata & 5.5 (0.6) & \nodata & $-$0.5 &  \\
 20.47 & +0.16 & 18 27 51  & $-$11 00  & 8 x 8 & Shell & I & 4.2 (0.1) & 2.7 (0.4) & 1.4 (0.1) & $-$0.3 & $-$0.5&  H \\
 21.04 & $-$0.47 &  18 31 12  & $-$10 47  & 9 x 7 & Shell & II & 2.3 (0.2) & 0.9 (0.5) & 0.5 (0.1) & $-$0.6 & $-$0.7&  \\
 21.56 & $-$0.10 & 18 30 50  & $-$10 09  & 5 x 5 & Partial Shell & II & 0.5 (0.1) & 0.3 (0.2) & 0.2 (0.1) & $-$0.4 & $-$0.6 &  H/X \\
\enddata
\tablenotetext{a} {``Partial shell'' morphology indicates that $\lesssim 70\%$ of a complete shell is evident in the 90cm image.}
\tablenotetext{b} {Statistical uncertainties are provided in
parenthesis, for the 90 and 21cm data they are calculated from ($\#$
independent beams)$^{0.5}\times 3\sigma$, while it is the larger of
$10\sigma$ or $10\%$ for the 11cm data.}
\tablenotetext{c} {Spectral indices between indicated wavelengths using $S_{\nu}\propto\nu^{\alpha}$; the uncertainty in $\alpha$ is typically the
larger of $\mid\alpha(90/20) - \alpha(90/11)\mid$ or 0.2.}
\tablenotetext{d} {H: also candidate in \citet{Helfand2005b}. O: previously reported in (ascending $\ell$ order) 
\citet{Yusef-Zadeh2000,Brogan2004}; \citet{Brogan2004,Trushkin1996,Brogan2004}; both \citet{Brogan2005} and \citet{Helfand2005a}; 
\citet{Odegard1986}. X: Possible {\em ASCA} X-ray counterpart in \citet{Sugizaki2001} or \citet{Bamba2003}.}
\end{deluxetable*}

The third criterion is useful in eliminating optically thick \HII\/
regions and thermal wind blown bubbles (WBB). Although the continuum
emission from most WBBs is thermal (i.e.\ no SN has happened inside
the bubble), they tend to have flat spectra and shell like
morphologies that are not easily distinguished from SNRs. However, WBB
(and \HII\/ regions) are also invariably surrounded by a shell of
bright 8 \mum\/ emission. While SNRs can also have associated mid-IR
emission \citep[e.g.][]{Reach2005}, this emission tends to be weak and
often undetectable at the sensitivity of {\em MSX}. Three-color images
of the W30 and W28 regions are shown in Figure~1a,b to demonstrate
both the high quality of the 90cm image and the candidate search
technique.

\begin{figure}
\epsscale{1.0}
\plotone{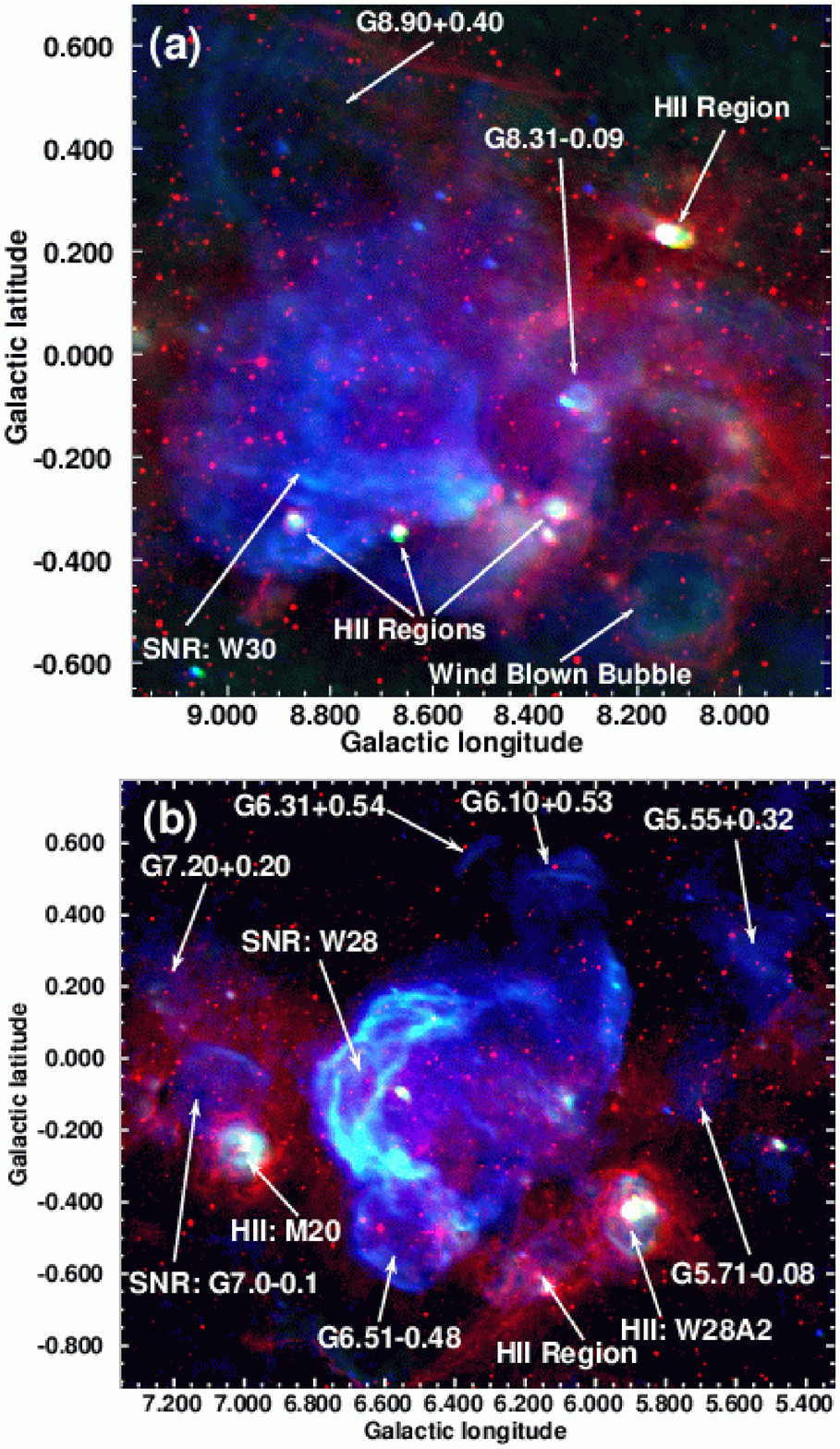}
\caption[]{Three color images with {\em blue}=VLA 90cm, {\em red}=MSX 8
\mum\/, and {\em green}=SGPS+VLA 20cm of the (a) W30 region and (b)
W28 region. New SNRs are indicated using Galactic coordinates; the
previously known SNRs, \HII\/ regions, and a WBB are also labeled.}
\end{figure}

Each candidate in Table~1 has been assigned a ``class'' of I, II, or
III: (I) we are very confident that the source is an SNR and all three
flux measurements are believed accurate; (II) we are fairly confident
that the source is an SNR, but the candidate is either confused with
thermal emission and especially the 11cm flux density is uncertain or
the source is large enough that the 20cm flux density could not be
estimated with the current data -- higher resolution and sensitivity
follow-up is desirable; (III) the candidate is coincident with
non-thermal emission but it is very faint, very confused, or does not
exhibit a typical shell type SNR morphology and future follow-up is
essential. There are 15 class-I, 16 class-II, and 4 class-III
candidates. Hereafter, all 35 sources will be designated as SNRs;
images of the new SNRs at 90cm are shown in Figure~2 (except for the
three described in Brogan et.\ al (2004)). Six of the new SNRs are
coincident with possible {\em ASCA} X-ray counterparts as indicated in
Table~1; no pulsars are coincident with any of the new SNRs based on
the \citet{Manchester2005} catalog.

\citet{Helfand2005b} recently reported the discovery of 30 SNR
candidates in the same region described in this work as part of their
larger Multi-Array Galactic Plane Imaging Survey (MAGPIS; the latitude
extent is $\mid b\mid<0.8\arcdeg$). Applying the criteria listed in
\S2.1, we find supporting evidence that 15 of the \citet{Helfand2005b}
sources are SNRs. The 15 candidates in common between the two surveys
are indicated in Table~1.  The \citet{Helfand2005b} 20cm
integrated flux densities are typically $\sim 4$ times higher than
those listed in Table~1 because their measurements were not restricted
to the candidate's spatial extent nor was background subtraction
performed.

\section{DISCUSSION}

Generally, the SNRs discovered in this survey are smaller and fainter
than those previously known in this region. The mean and median
diameters of the new SNRs are $\sim 12\arcmin$ and $8\arcmin$. The
SNR candidates have 1~GHz surface brightnesses in the range
$\Sigma_{\rm 1~GHz}=(1-15)\times 10^{-21}$
W~m$^{-2}$~Hz$^{-1}$~sr$^{-1}$. This result suggests that the Bonn
11cm completeness limit estimated by \citet{Green2004} of $\sim
10^{-20}$ W~m$^{-2}$~Hz$^{-1}$~sr$^{-1}$ is accurate and that the
current survey is $\sim 10$ times deeper.

This limited 90cm survey of 42.5~deg${}^2$ has produced 
a $\sim 15\%$ increase in the total number of known Galactic SNRs. The
number of identified remnants within the survey boundaries (previously 19)
has increased by nearly a factor of 3 to 54.  Based on completeness
considerations and empirical evidence, respectively,
\citet{Helfand1989} and \citet{Brogan2004} estimate that there should
be between 65 to 85 SNRs in this region, in reasonable agreement with
the observed number, given that we are insensitive to the smallest
($\la 2.5\arcmin$) and largest ($\ga 0.8\arcdeg$) size scale remnants,
as well as Crab like remnants lacking radio shells.

To estimate the implications of these new SNR discoveries on the total
number of Galactic SNRs we assume (1) the current SNR catalog
\citep{Green2004} is essentially complete for $\mid b\mid >
1.25\arcdeg$ and for $\mid\ell\mid >50\arcdeg$; (2) that the Bonn 11cm
survey in the 1st quadrant and the MOST 35cm survey in the 4th
quadrant have similar levels of completeness
\citep[e.g.][]{Green2004}; and (3) the number of known SNRs between
$\mid b\mid< 1.25\arcdeg$ and $\mid\ell\mid <50\arcdeg$ (115) should
be scaled by a factor of three (i.e.\ the increase achieved in our
limited survey). The $\ell=50\arcdeg$ cutoff is guided by the
longitude extent of the inner Galactic disk. These assumptions result
in a prediction of 460 for the number of Galactic SNRs, still a factor
of two less than the expected $\sim 1,000$ SNRs (see \S 1).  

However, five of the new SNRs are in close proximity to (or coincident
with) the two largest previously known SNRs in the survey region: W28
and W30 (Figs.~1a,b), but appear distinct from them. Although we do not
currently know if the new SNRs are physically close to the two known
SNRs (along the line of sight), such a result would be unsurprising
since high mass stars form in clustered environments. A new SNR has
also been proposed to lie along the line of sight to the Vela SNR
\citep[i.e.\ G266.2--1.2][]{Aschenbach1998}. Such superpositions
throughout the plane may also explain some fraction of the ``missing''
remnants. The assumption of completeness for $\mid\ell\mid >50\arcdeg$
is also probably flawed as a few SNRs continue to be discovered in the
outer Galaxy \citep[e.g.][]{Kothes2001}.
 
Overall, these results suggest that the ``missing SNRs'' problem can
be attributed to selection effects and not our understanding of SN
rates.  Future instruments like the EVLA, LWA, ATA, and SKA will allow
even deeper low frequency, high dynamic range surveys which will likely
discover the remaining shortfall of Galactic SNRs.

\acknowledgments

The National Radio Astronomy Observatory operates the Very Large Array
and is a facility of the National Science Foundation operated under a
cooperative agreement by Associated Universities, Inc.  This research
made use of data products from the Midcourse Space Experiment. Basic
research in radio astronomy at the NRL is supported by the Office of
Naval Research. We would also like to thank the SGPS team for their
hard work in acquiring the ATCA 20cm data.

\begin{figure}
\epsscale{1.0}
\plotone{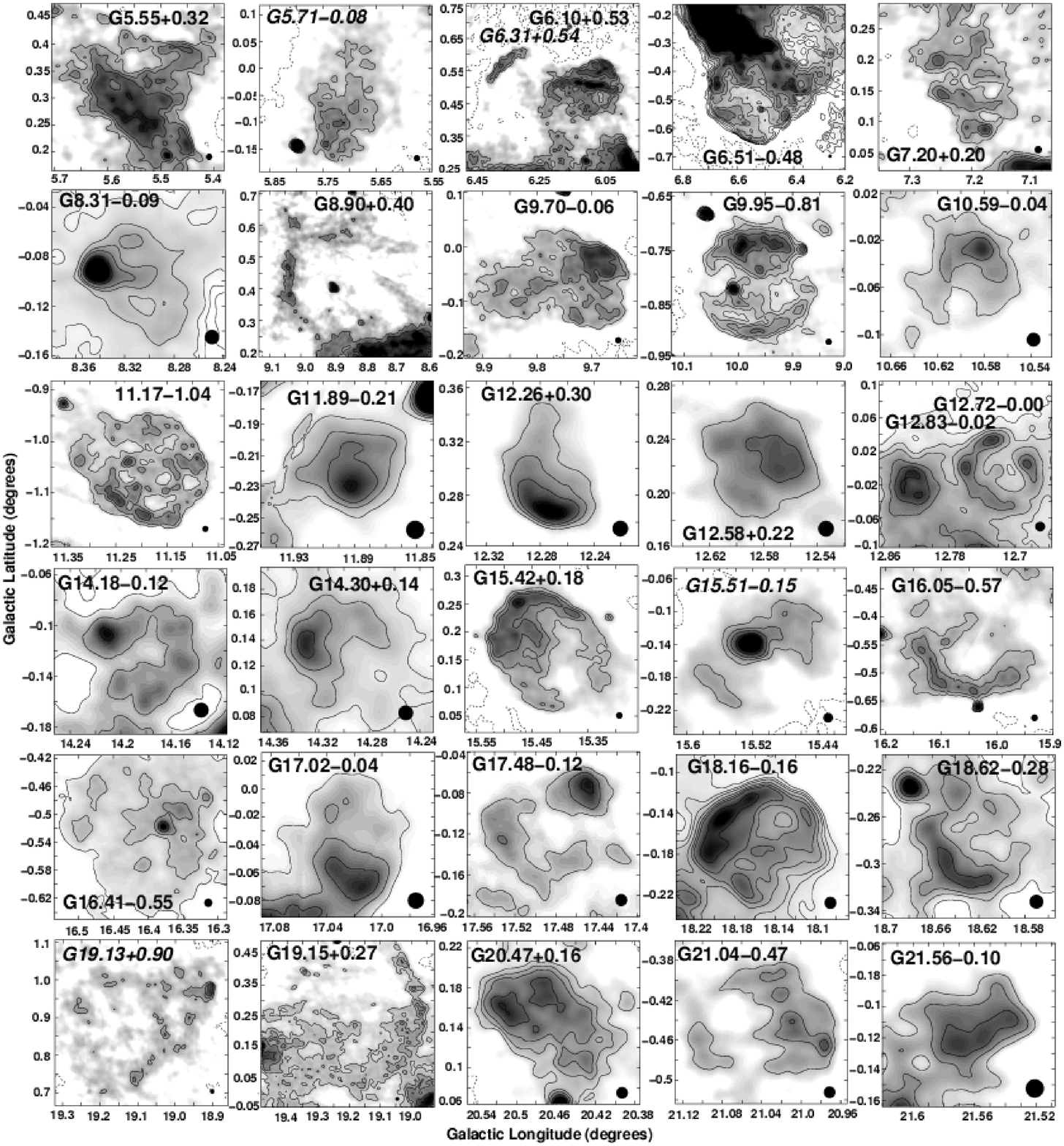}
\caption[]{VLA 90cm images of the newly discovered SNRs (excluding
three shown in \citet{Brogan2004}). The contour levels on
each image are -25, 25, 37.5, 50, 75, 100, \& 125 \mjb\/. For
latitudes $\mid b\mid\lesssim 0.5\arcdeg$, 25 \mjb\/ is approximately
$5\sigma$, while it is $\sim 3\sigma$ at higher latitudes. The
resolution of these images is $42\arcsec$; the beam size is shown in
the lower right of each panel for comparison. Class III sources are labeled in 
italics.}
\end{figure}

\end{document}